# The challenge of attosecond pulse metrology

J. E. Kruse[1], P. Tzallas[1*], E. Skantzakis[1], C. Kalpouzos[1], G. D. Tsakiris[3] and D. Charalambidis[1,2]

[1]*Foundation for Research and Technology—Hellas, Institute of Electronic Structure and Laser, PO Box 1527, GR-711 10 Heraklion, Crete, Greece*

[2]*Department of Physics, University of Crete, PO Box 2208, GR-71003 Heraklion, Crete, Greece*

[3]*Max-Planck-Institut für Quantenoptik, Hans-Kopfermann-Str. 1, D-85748 Garching, Germany*

*Corresponding author e-mail address: ptzallas@iesl.forth.gr

The two basic approaches underlying the metrology of attosecond pulse trains are compared, i.e. the 2nd order Intensity Volume Autocorrelation and the Resolution of Attosecond Beating by Interference of Two photon Transitions (RABITT). They give rather dissimilar results with respect to the measured pulse durations. It is concluded that RABITT may underestimate the duration due to variations of the driving intensity, but in conjunction with theory, allows an estimation of the relative contributions of two different electron trajectories to the extreme-ultraviolet emission.

One of the main barriers that *attoscience* [1] had to overcome in its early stages was the development of reliable temporal characterization techniques. Established extreme-ultraviolet (XUV) attosecond (*asec*) pulse metrology is widely based on two approaches: I) the cross-correlation of the driving IR with the generated XUV fields [2, 3], particularly favorable for the characterization of moderate intensity *asec* pulses, and II) the 2$^{nd}$ order Intensity Volume Autocorrelation (2-IVAC) approach [4] requiring high XUV intensities, indispensable for the observation of a non-linear process induced solely by the XUV radiation to be characterized [4, 5]. High XUV intensities are currently provided only by *asec* pulse trains, generated using many cycle high peak power lasers. Further, less frequently applied approaches include SPIDER [6] and the in-situ method [7].

In the metrology of *asec* pulse trains, the pioneering method of RABITT underlies almost all cross-correlation based techniques [2, 3]. Relative harmonic phases are extracted from the RABITT traces, and from those the *asec* train is reconstructed. The alternative method of the 2-IVAC relies on a two-XUV- photon atomic ionization by two replicas of the XUV pulse train. 2-IVAC [4] does not permit reconstruction of the XUV waveform, but reveals to a satisfactory degree of accuracy the pulse duration. Despite long standing debates on the results and the applicability of the two techniques, *asec* pulse metrology lacks direct comparative measurements between them. This letter is meant to fill this gap and reach conclusions on unexplored possibilities and limitations in attosecond metrology.

Odd harmonics of a Ti:Sapphire laser have been generated by focusing 55 fs laser pulses, at intensities $I_L \sim 10^{14}$ W/cm$^2$ into a xenon gas jet. The temporal characteristics of the emitted harmonic superposition have been measured by both the 2-IVAC and the RABITT technique. Care was taken that the generation conditions were identical when applying both methods. The set-up used is described in detail in ref. [8]. In both, 2-IVAC and RABITT an annular laser beam of 1.9 cm outer diameter and 13 mJ/pulse was used. For RABITT, a mask with a hole in the center was used to create a 1.5 mm diameter IR beam in the center of the annular beam in a similar manner as in ref. [2] with a pair of 3 mm thick plates to control of the temporal delay between the central and annular beam. After the xenon jet a silicon wafer was placed at the fundamental's Brewster angle of 75° to reduce the IR and to reflect the harmonics [9] towards the detection area. For RABITT, a λ/2 waveplate was used to rotate the polarization of the laser by a very small angle (<3 deg) in order for some of the central IR dressing-beam to be reflected towards the detection area. Measurements have been performed for three different positions of the laser focus with respect to the position of the gas jet: I) $z_{BJ}$ = - 0.43$b$ (laser focus before jet), II) $z_{OJ}$ = 0 (laser focus at jet) and III) $z_{AJ}$ = + 0.28$b$ (laser focus after jet). $b$=17.6 ± 0.9 cm is the confocal parameter, assuming Gaussian beam geometry.

According to the three step model [10, 11], two trajectories of the electron, driven by the IR field, contribute to each emitted harmonic. Due to their different lengths they are termed the "short" and "long" trajectory. It is known from modeling [12, 13] that phase matching favors in case I) the contribution of the "short" trajectory, in case II) both and in case III) the contribution of the "long" trajectory to the harmonic emission. Here experiments were performed for all three focus positions, precisely in order to investigate the phase matching behavior of the harmonic emission associated with each trajectory and their relative contributions.

An iris put downstream the XUV beam was selecting the central part of the XUV beam, to which the short trajectory has an enhanced contribution [13-16], in order to ensure experimental conditions similar to those of all previous 2-IVAC and some RABITT experiments, but not the very first one [2].

In 2-IVAC, the radiation reflected from the wafer was passing through a 150 nm thick indium filter to select the 9[th]-15[th] harmonic and also to block any residual IR. This beam was then focused by a split spherical gold mirror of 5 cm focal length into a helium pulsed gas jet. The relative field amplitudes of the harmonics in the interaction region were measured to be 100:40:30:25, for the 9$^{th}$, 11$^{th}$, 13$^{th}$ and 15$^{th}$ harmonic, respectively. The ionization products were recorded by a μ-metal-shielded time-of-flight (TOF) spectrometer.

Figure 1a depicts the harmonic yield as a function of the laser focus position. Figures 1b-d shows recorded 2-IVAC traces for the three focus positions. A clear *asec* pulse structure has been observed only at $z_{BJ}$ with average pulse duration of 660 ± 50 asec. This is in agreement with theoretical predictions [12, 13]. For the other positions some irregular signal modulation is present, indicating absence of a clear *asec* pulse train.

In RABITT, the resulting two-color beam was reflected by only one half of the split mirror and focused into an argon pulsed gas jet. The two-color (IR+XUV) photoelectron (PE) spectra have been recorded as a function of the delay τ between the generated XUV and the IR dressing-beam by rotating one plate over 18 fs near zero delay. The PE spectra show a series of 11$^{th}$-17$^{th}$ harmonic single-photon ionization peaks and between them two-photon ionization (IR+XUV) "sideband" peaks S12-S16.

Figure 2 shows RABITT traces recorded for the three focus positions from which relative harmonic phases have been extracted and the harmonic superposition has been reconstructed. In contrast to the results of the 2-IVAC, for all three focus positions, the waveform exhibits pronounced *asec* pulse structure. Pulse durations of 390 *asec* are found for $z_{BJ}$ and $z_{OJ}$, and 630 *asec* for $z_{AJ}$. For a direct comparison between the two techniques, the durations have been corrected according to the dispersion of the indium filter [17], used for the 2-IVAC. Then the durations become 350 *asec*, 600 *asec* and 1020 *asec* for the positions $z_{BJ}$, $z_{OJ}$ and $z_{AJ}$, respectively. This result is in contradiction with both the experimental findings of the 2-IVAC, as well as the theoretical predictions [12, 13].

Figure 3a shows the extracted relative harmonic phases and Fig. 3b the corresponding emission time ($t_e$) and time shift (Δ$t_e$) for the three focus positions. While the focus position moves past the gas jet, the harmonic chirp is passes from positive to negative values. This is why for $z_{BJ}$, $z_{OJ}$ the same duration but opposite pulse asymmetry has been measured. This result is compatible with the presence of two electron trajectories at all three positions, but with different contributions to the emission at each position. Further evidence of the presence of two trajectories is given by the periodic modulation of the harmonic yield as a function of the laser intensity observed at different focus positions, which will be the subject of a forthcoming publication [18]. Similar behavior is also observable in a careful evaluation of a line out of Fig. 14a, c of ref. 19. In particular, the presence of the long trajectory is detrimental to attosecond localization, since its $t_e$ and $\varphi_q$ strongly depend on the driving laser intensity. It has been shown [10, 12-14] that laser intensity modulations are broadening the XUV pulses. RABITT though measures phases as a spatiotemporal and shot-to-shot average $<\varphi_q>$, which artificially leads to a very short reconstructed pulse, when both trajectories are contributing, resulting in compensating negative and positive chirp values in the averaged phases. On the contrary, the 2-IVAC measures a pulse duration, which is between the average duration of the waveforms emitted at different driving intensities and the duration of the superposition of these waveforms. This is closer to the pulse duration relevant to time resolved applications of the *asec* pulses.

The RABITT traces of fig. 5 were calculated as follows: considering both trajectories contributing to the harmonic generation process, the modulation of the $(q+1)^{th}$ sideband signal $S_{q+1}$ with the delay $\tau$ reads

$$S_{q+1} \propto \left(I_q^S I_{q+2}^S\right)^{1/2} \cos(2\omega_L\tau + \Delta\varphi_{q,q+2}^S) + \left(I_q^L I_{q+2}^L\right)^{1/2} \cos(2\omega_L\tau + \Delta\varphi_{q,q+2}^L) + \\ + 2\left(I_q^S I_{q+2}^S I_q^L I_{q+2}^L\right)^{1/4} \cos(A_{q+1})\cos(2\omega_L\tau + \frac{\Delta\varphi_{q,q+2}^S + \Delta\varphi_{q,q+2}^L}{2}), \quad (1)$$

where $A_{q+1} = \frac{1}{2}(\Delta\varphi_q^{SL} + \Delta\varphi_{q+2}^{SL})$, $\Delta\varphi_{q,q+2}^S = \varphi_q^S - \varphi_{q+2}^S$, $\Delta\varphi_{q,q+2}^L = \varphi_q^L - \varphi_{q+2}^L$, $\Delta\varphi_q^{SL} = \varphi_q^S - \varphi_q^L$, $\Delta\varphi_{q+2}^{SL} = \varphi_{q+2}^S - \varphi_{q+2}^L$. Here, $I_q^S$ and $I_q^L$ are the normalized intensities ($I_q^S + I_q^L = 1$) of the harmonic order $q$ generated by the short and long trajectories, respectively. $\varphi_q^S$, $\varphi_q^L$ are the phases of harmonics generated by electrons from the short and long trajectory, respectively, and $\omega_L$ is the fundamental laser frequency. In eq. 1 the atomic phase shift ($\Delta\varphi_{atom}$) is assumed to be zero due to its negligible influence on the sideband signal. Further it is assumed that the normalized contribution of the short trajectory to the $S_{q+1}$ sideband is $\mu_{q+1}$ ($0 \leq \mu_{q+1} \leq 1$) and that of the long trajectory ($1-\mu_{q+1}$). The $\mu_{q+1}$ is treated as a "mean" of the normalized contributions $\mu_q$ and $\mu_{q+2}$ of the short trajectory to the harmonics $q$ and $q+2$, which are approximated to be equal i.e. $\mu_q \approx \mu_{q+1} \approx \mu_{q+2}$. In this approximation $E_q^S \approx \mu_{q+1} E_q$, $E_{q+2}^S \approx \mu_{q+1} E_{q+2}$ and $E_q^L \approx (1-\mu_{q+1}) E_q$, with $E_q$ and $E_{q+2}$ being the corresponding harmonic fields. Then for each sideband the "mean" $\mu_{q+1}$ may have a different value. Although the approach is not rigorous, the above approximation is justified for a smooth variation of $\mu$ along the spectrum, allowing a reasonable estimation of the relative contribution of the two trajectories.

In Fig. 4 the phases of the harmonics were calculated from the quantum mechanical version of the three step model [10]. Propagation effects in the harmonic generation medium were not taken into account. A 15% intensity variation around $6 \times 10^{13}$ W/cm$^2$ was included in the calculation. At this intensity the 19$^{th}$ harmonic is in the cut-off region.

The colour plots of Fig. 4a-c depict the calculated modulation of the signal of S12-S16 as a function of the delay $\tau$ and the relative contributions of $I_q^S$ and $I_q^L$. It is apparent, that the phase relation and thus the pulse durations, that would be extracted from traces of Fig. 4a-c, strongly depend on the chosen $I_q^S / I_q^L$ ratio. As can be seen from Fig. 4, for $I_q^S = I_q^L$ the harmonic chirp is almost completely vanishing due to the compensating contributions of the negative and positive chirp. In this case the pulse reconstruction will result in an almost FTL attosecond pulse train.

To elucidate the above, the line in Fig. 4d shows the calculated superposition of several XUV waveforms resulting from different driving intensities at equal contributions from the short and long trajectories. The green filled area in Fig. 4d shows the corresponding waveforms reconstructed using the averaged phases that RABITT would measure. The difference between the pulse trains of Fig. 4d is glaring. From the above it becomes apparent that RABITT can reveal neither the real temporal pulse profile nor the pulse duration, unless only one trajectory is present, which is certain only for the cut-off harmonics. However, in conjunction with the three step model it can be used to estimate the relative contributions of the two trajectories to the emitted XUV field.

Comparing the measured with calculated RABITT traces of Fig. 4a-c, we find that the contribution of the long trajectory increases with harmonic order and $z$ position. The contribution of the long trajectory is found for S12, S14 and S16 to be 23%, 33% and 55% respectively at $z_{BJ}$, 23%, 40% and 65% at $z_{OJ}$ and 23%, 45% and 95% at $z_{AJ}$. Note that the traces in Fig. 4a-c are very sensitive to the IR intensity. Thus the accuracy of the extracted contributions is determined by the accuracy at which the IR intensity is known.

The results of this first comparative study of the 2-IVAC versus the RABITT technique evinces, that the latter is apt to reveal the temporal profile of the waveform only if one of the two electron trajectories, and in particular the long one, is fully eliminated - an experimentally cumbersome situation for plateau harmonics. However, through RABITT the relative contributions of the two electron trajectories in the harmonic generation process may be estimated. Thus, the two techniques are complementary in attosecond pulse metrology, RABITT shading light on the dynamics of the emission process, while 2-IVAC measures the pulse duration.

**Acknowledgments**

This work is supported in part by the European Commission Human Potential Program under contract MTKD-CT-2004-517145 (X-HOMES); the Ultraviolet Laser Facility (ULF) operating at FORTH-IESL (contract no. HPRI-CT-2001-00139), the ELI research infrastructure preparatory phase program, and the Marie Curie Research and Training Network Contr. Nr. MRTN-CT-2003-505138 (XTRA).

**Figure Captions**

**Figure 1.** (a) Harmonic signal as a function of the focus position relative to the xenon gas jet. The arrows show the positions $z_{BJ}$, $z_{OJ}$ and $z_{AJ}$, where the 2-IVAC and RABITT measurements were performed. (b)–(d) 2-IVAC traces. The gray dots are the measured data. The gray circles correspond to a 10 point running average. The purple line in (b) is a 12-peak sum of Gaussians fit to the raw data. The green area in (b) is one of the 12 Gaussian pulses of the fitted function.

**Figure 2.** (Left panel) RABITT traces measured at the three focus positions normalized to the corresponding total signal. The gray points are the measured data. The red, yellow-filled circles correspond to a running average over 15 points for the sideband traces and over 40 points for the total signal (Total). The purple lines on the sideband traces are sinusoidal fits to the raw data over 13 oscillations. The purple lines on the total signal traces are sinusoidal fits to the raw data over 6 $T_L$ oscillations. (Right panel) Reconstructed pulse trains.

**Figure 3.** (a) Harmonic phases $\varphi_q$ and (b) electron return times $t_e$ as a function of the harmonic order q obtained from the RABITT traces of Fig. 2. The purple lines in (a) are quadratic fits to the data, in (b) linear fits to the data. The $t_e$ is set to zero for the cut-off harmonics. The gray filled circle corresponds to the position of the 19$^{th}$ harmonic which is assumed to belong to the cut-off region.

**Figure 4.** (a)–(c) Calculated RABITT traces for different relative contributions of the short and long trajectories. The solid yellow, the dashed blue, and the solid black lines match the measured RABITT traces at $z_{BJ}$, $z_{OJ}$ and $z_{AJ}$, respectively. (d) The orange line is the superposition of several attosecond pulse trains calculated using equal contributions of the short and long trajectories. The green area shows the attosecond pulse train, that RABITT would result in for the above superposition.

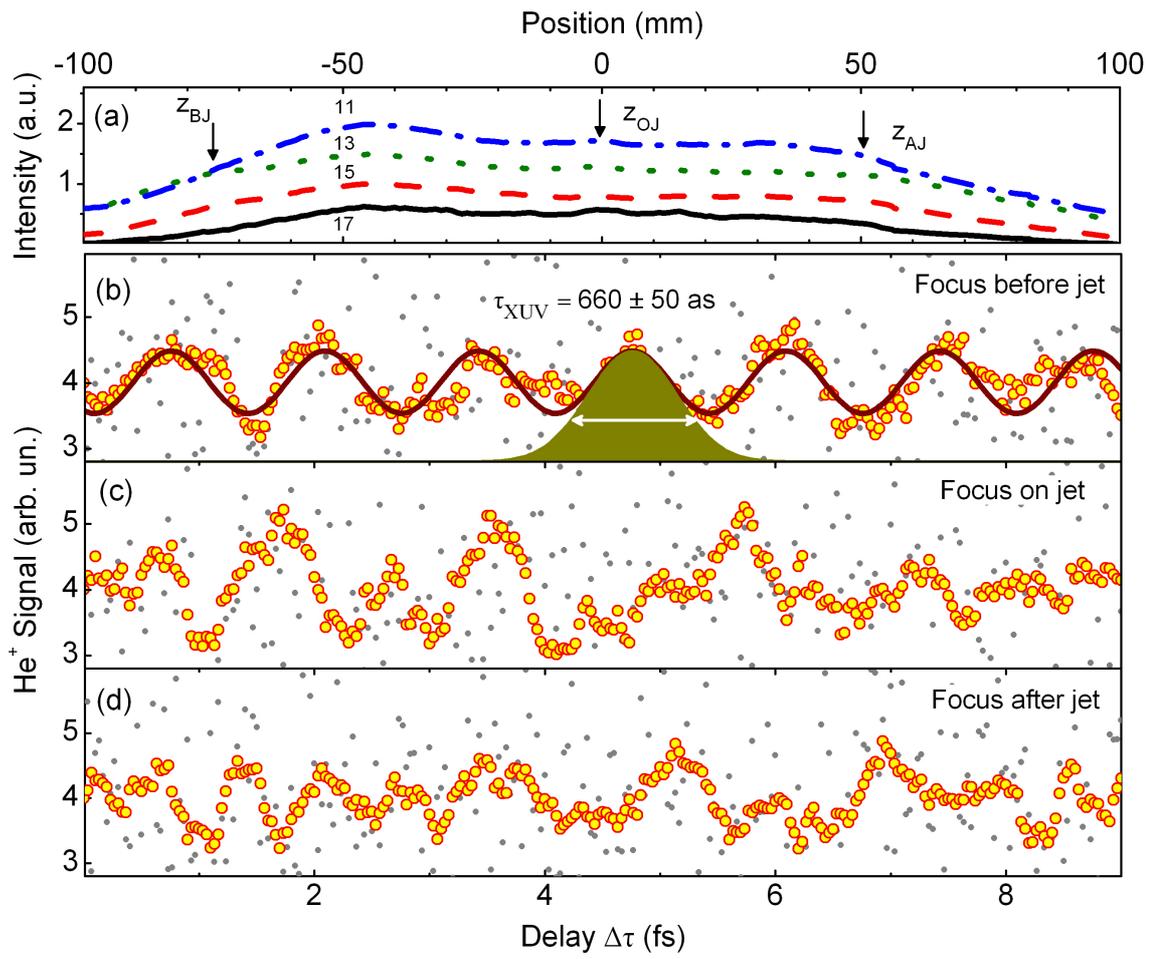

Fig.1

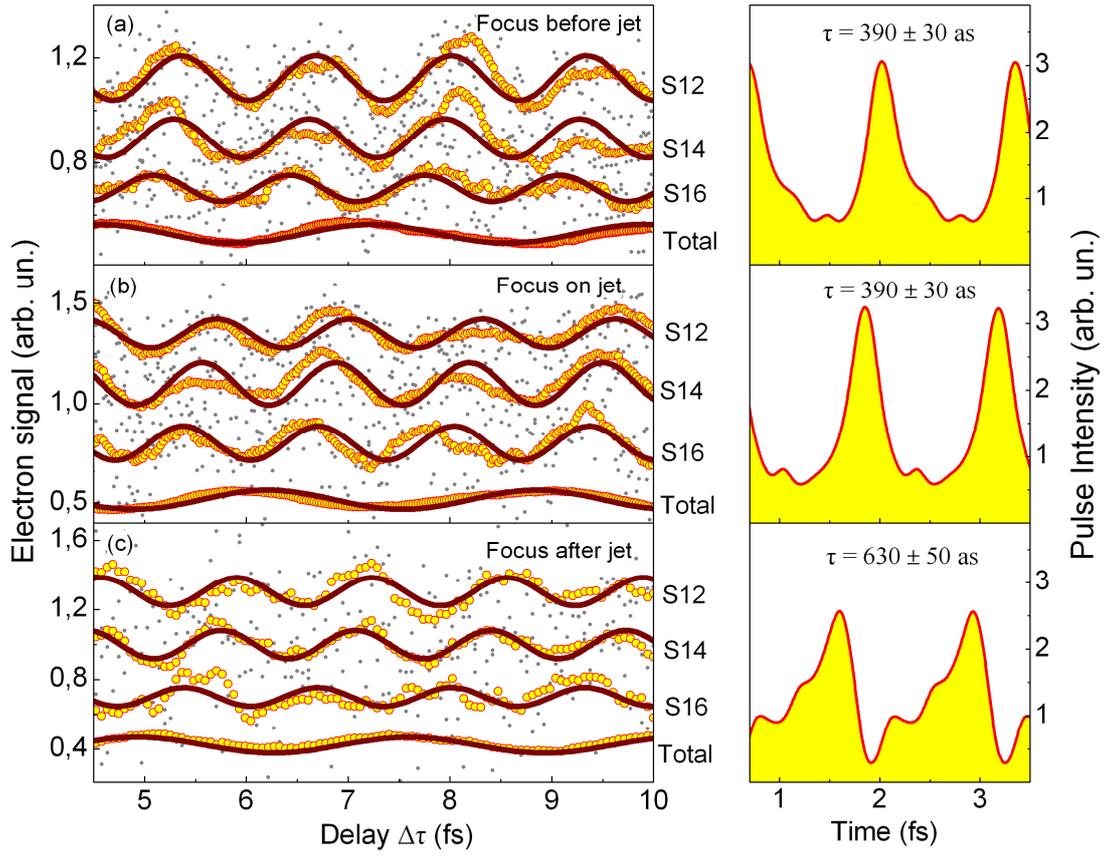

**Fig. 2**

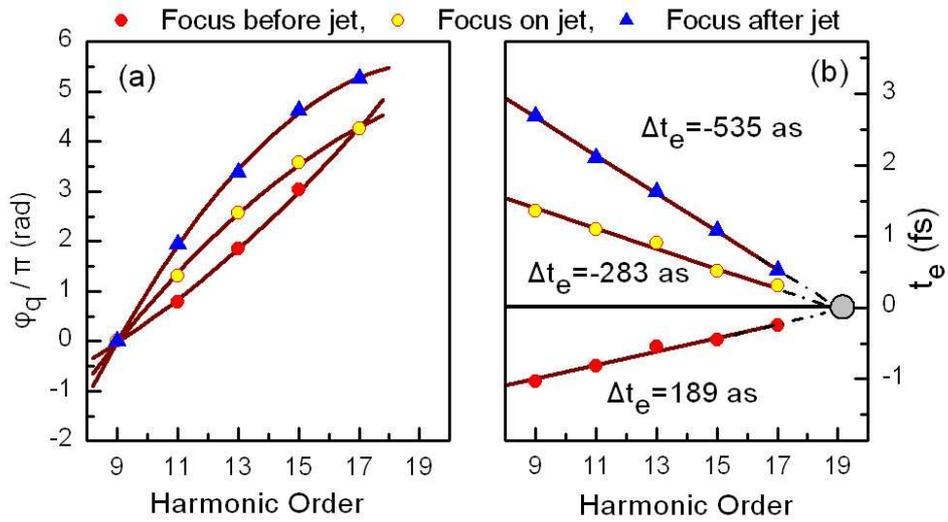

**Fig.3**

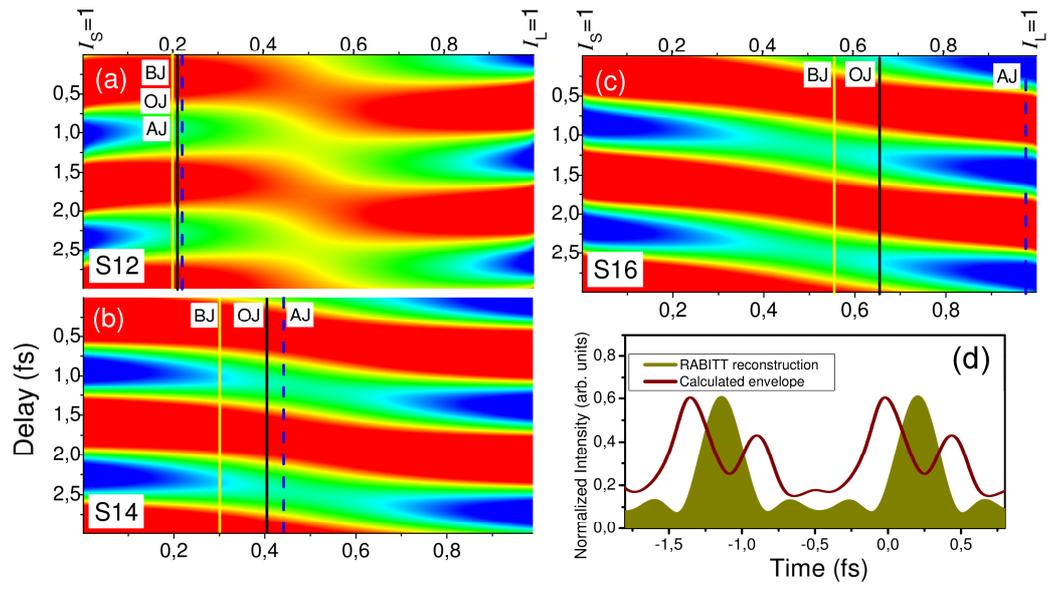

Fig.4